\documentclass[3p]{elsarticle}
\usepackage{lineno,hyperref}
\usepackage{graphicx}
\usepackage{amsmath}
\usepackage{color}
\usepackage[normalem]{ulem}

\journal{arXiv}
\begin{document}

\begin{frontmatter}

\title{Implied and Realized Volatility: A Study of the Ratio Distribution}

\author[mymainaddress]{M. Dashti Moghaddam}
\author[mymainaddress]{R. A. Serota\fnref{myfootnote}}
\fntext[myfootnote]{serota@ucmail.uc.edu}

\address[mymainaddress]{Department of Physics, University of Cincinnati, Cincinnati, Ohio 45221-0011}

\begin{abstract}
We analyze correlations between squared volatility indices, VIX and VXO, and realized variances -- the known one, for the current month, and the predicted one, for the following month. We show that the ratio of the two is best fitted by a Beta Prime distribution, whose shape parameters depend strongly on which of the two months is used.
\end{abstract}


\end{frontmatter}

\section{Introduction}

In a previous study \cite{dashti2018implied}, we introduced the ratio of realized variance to implied variance, represented by squared volatility indices VIX or VXO, as a measure of their correlations. We pointed out that the realized variance is calculated for trading dates while the implied variance covers every day, so one of them needs rescaled for a proper comparison. We argued that studying the distribution of the ratios produces a deeper insight into these correlations than a simple regression analysis \cite{christensen1998relation}, which arrives at an obvious conclusion that VIX/VXO are a slightly better predictor of the future realized volatility (RV) than the past RV since it builds on the latter with the benefit of additional information. 

In \cite{dashti2018implied} we concluded that the ratio of the actual realized variance $RV^2$ to $VIX^2$ and $VXO^2$, that is to its predicted values, was best described by the fat-tailed inverse Gamma (IGa) distribution and its inverse by the Gamma (Ga) distribution.We speculated that this is due to unanticipated spikes of realized volatility. In this paper we show that a Beta Prime (BP) distribution provides a better fit both for the ratio and its inverse. For the former, the exponent of the  power dependence for small values of the ratio is very large, which mimics exponential behavior of IGa. For the latter, the exponent of the power-law tails is very large, which mimics exponential decay of Ga.

We also concluded in \cite{dashti2018implied} that the ratio of $RV^2$ of the preceding month to $VIX^2$ and $VXO^2$ was best described by the lognormal (LN) distribution. Its inverse was also best described by the LN with similar parameters. We argued that while the spikes in the past realized volatility lead to spikes of implied volatility, there was enough uncertainty for the ratio to have heavy tails. In this paper we show that BP, with similar parameters for the ratio and its inverse, provides a better fit for both. The exponents of the power law for both small values and the fat tails are large, so that BP mimics the LN behavior. Additionally, for both BP and LN the distribution of the inverse variable is BP and LN respectively as well. 

As a reminder to the reader, PDF of BP distribution is given by
\begin{equation}
BP(p, q, \beta; x) = \frac{(1+\frac{x}{\beta})^{-p-q}(\frac{x}{\beta})^{p-1}}{\beta \space B[p,q]}
\label{BetaPrimePDF}
\end{equation}
where $\beta$ is a scale parameter, $p$ and $q$ are shape parameters and $B[p,q]$ is the Beta function; $BP \propto x^{p-1}$ for $x \ll \beta$ and $BP \propto x^{-q-1}$ for $x \gg \beta$.

This paper is organized as follows. In Section \ref{EO} we summarize empirical observations regarding distributions of $RV^2$ and volatility indices $VIX^2$ and $VXO^2$ and their ratios. In Section \ref{RV} we show the results of statistical fits of the ratios. In Section \ref{Correlations} we summarize correlations between the quantities and their ratios.

\section{Empirical Observations \label{EO}}

In \cite{dashti2018implied} we presented empirical distributions (PDF) of $RV^2$ vis-a-vis $VIX^2$ and $VXO^2$, as well as the ratios, and in \cite{dashti2018realized} we will have described their fits. We observe the following features, in agreement with \cite{russon2017nonlinear} (see also \cite{vodenska2013understanding,kownatzki2016howgood}):

\begin{itemize}
  \item $VIX^2$ and $VXO^2$ have lower high-volatility probabilities relative to $RV^2$, including shorter fat tails, indicating that volatility indices do not predict accurately large values of RV, including the largest volatility spikes. In other word, volatility indices underestimate future large RV. 
   \item $VIX^2$ and $VXO^2$ have higher mid-volatility probabilities relative to $RV^2$, indicating that volatility indices overestimate future mid-level RV.
  \item $VIX^2$ and $VXO^2$ have lower low-volatility probabilities relative to $RV^2$, indicating that volatility indices underestimate future low RV.
\end{itemize}

For the distributions of the ratios $RV^2/VIX^2$, $RV^2/VXO^2$, it is important to notice that, since realized and implied volatilities are correlated, we cannot construct them simply as the quotient distributions of two independent variable. We observe the following:

\subsection{Predicted Month}

For the month predicted by the volatility indices

\begin{itemize}
  \item The distributions have fat tails, indicating again that VIX and VXO underestimate future values of RV, in particular volatility spikes.
  \item Very small ratios are suppressed, as manifested by a very large power exponent, indicating that it is rare that RV is considerably smaller than the one predicted by the volatility indices. 
  \item The tail exponents of the ratio distributions is larger than that of either $RV^2$ or $VIX^2$ and $VXO^2$, pointing to that for the $RV^2$ values taken from the tails, the values of $VIX^2$, $VXO^2$ are also more likely to come from the tails. 
\end{itemize}

\subsection{Preeceding Month}

For RV of the preceding month:

\begin{itemize}
  \item The tails of the distributions are much shorter than those for the predicted month, reflecting the fact that volatility indices account for past RV.
  \item The tail exponents of the distributions are almost identical to those of their inverse, $VIX^2/RV^2$ and $VXO^2/RV^2$ distributions, indicating, as above, strong correlations. 
\end{itemize}

For the ratio distribution of $RV^2$ of the predicted (next) month to $RV^2$ of the preceding month (see below), we observe that 

\begin{itemize}
  \item The exponent of the fat tail is smaller than those of the $RV^2/VIX^2$ and $RV^2/VXO^2$ distributions, that is the tails are longer. 
  \item The power-law exponent at very small ratios is much smaller for this distribution than for $RV^2/VIX^2$ and $RV^2/VXO^2$, that is those ratios are far less suppressed. 
\end{itemize}

By both measures, VIX and VXO are better predictors of the future RV than the past RV. 

\section{Statistical Fits of Ratio Distributions \label{RV}} 

Below, figures show the plots of the ratios and their distribution fits and tables contain parameters of the distribution and KS statistics. The two novel elements here, relative to \cite{dashti2018implied}, is the inclusion of BP in the fits of $RV^2/VIX^2$ and $RV^2/VXO^2$ distributions and the fits of the ratio distribution of $RV^2$ of the predicted (next) month to $RV^2$ of the preceding month (and its inverse).

\clearpage
\subsection{Predicted Month}

\begin{figure}[!htbp]
\centering
\begin{tabular}{cc}
\includegraphics[width = 0.49 \textwidth]{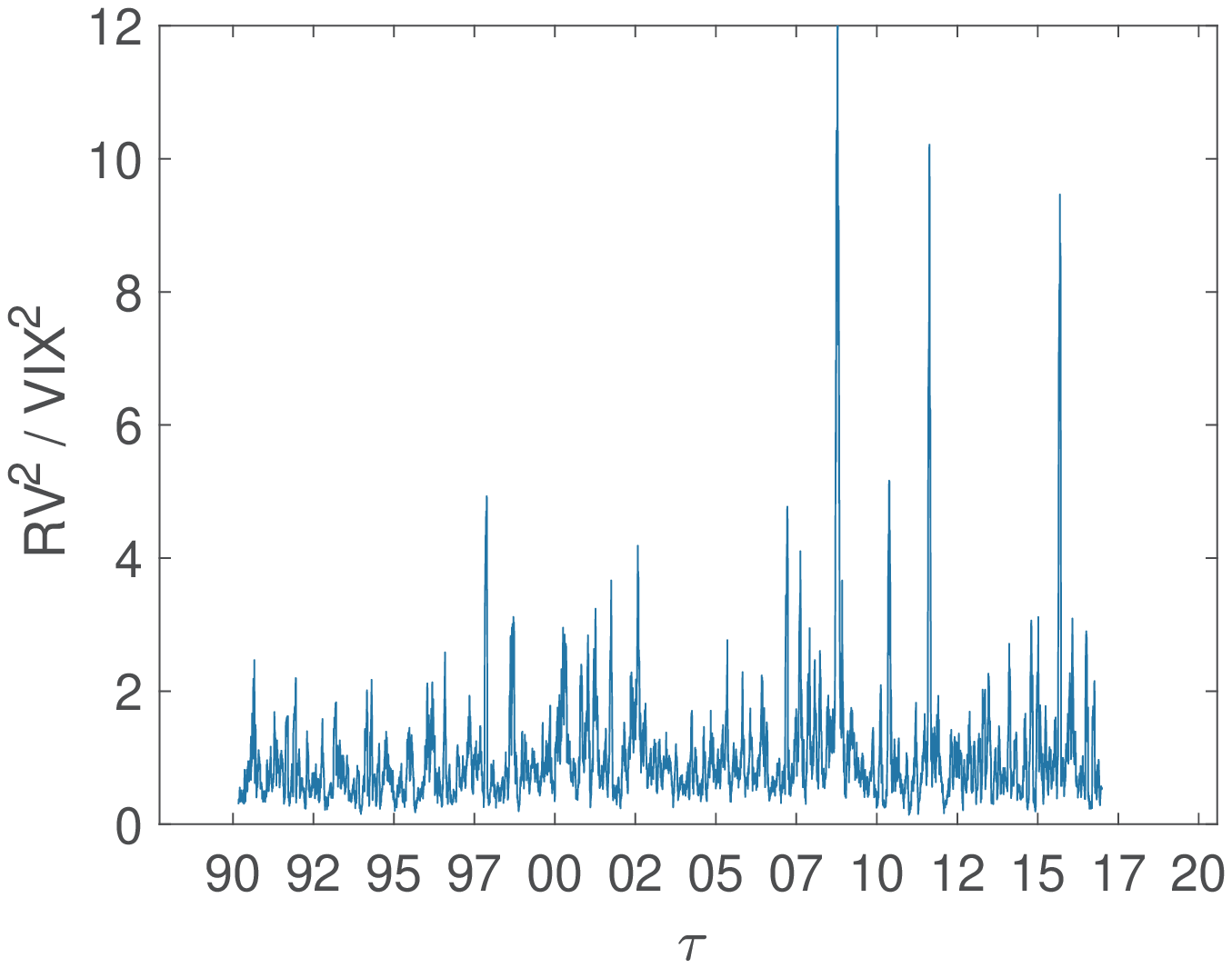}
\includegraphics[width = 0.49 \textwidth]{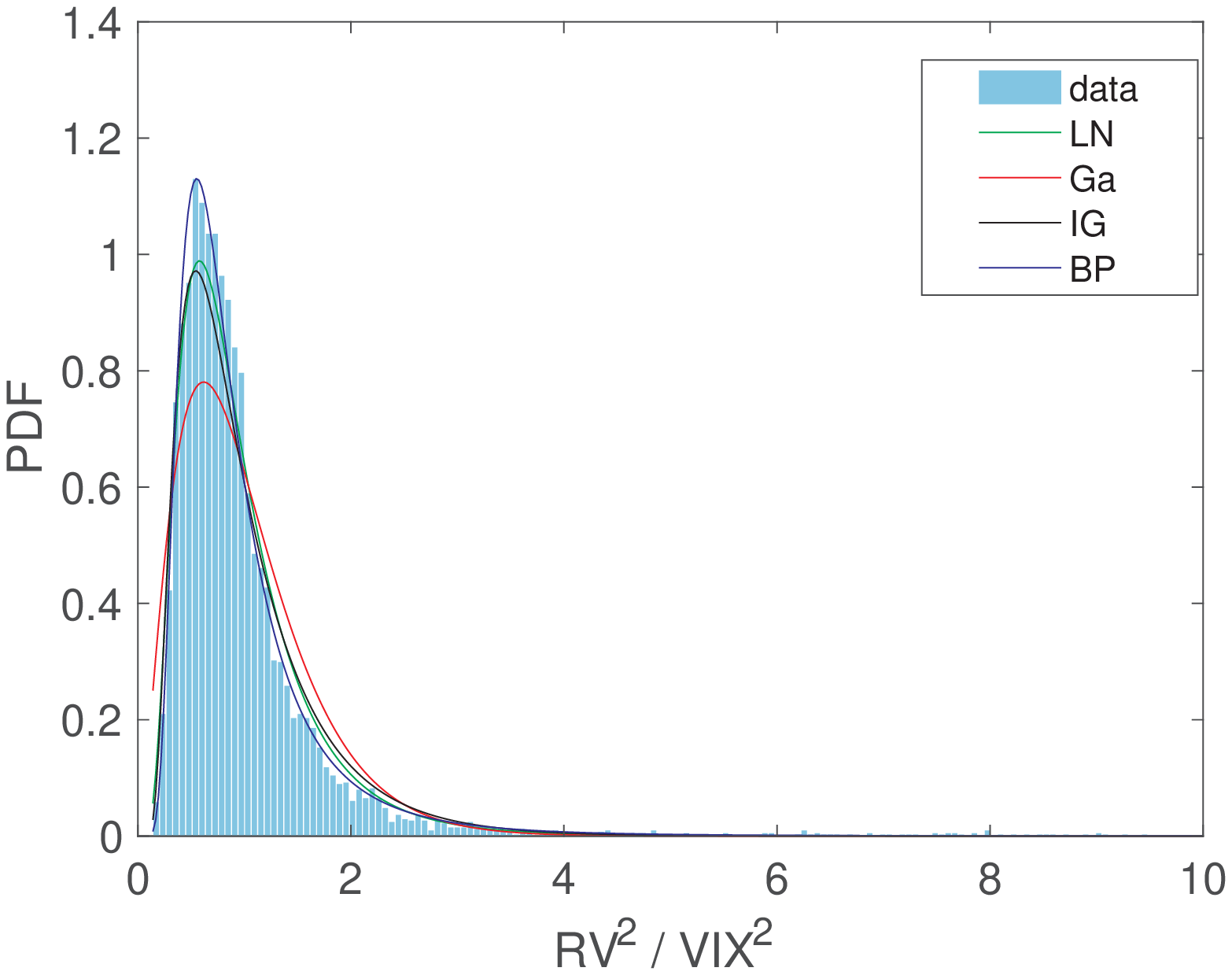}
\end{tabular}
\caption{$\mathrm{RV}^2 / \mathrm{VIX}^2$, from Jan 2nd, 1990 to Dec 30th, 2016.}
\label{RVOverVIXListSRV2OverVIX21990}
\end{figure}

\begin{figure}[!htbp]
\centering
\begin{tabular}{cc}
\includegraphics[width = 0.49 \textwidth]{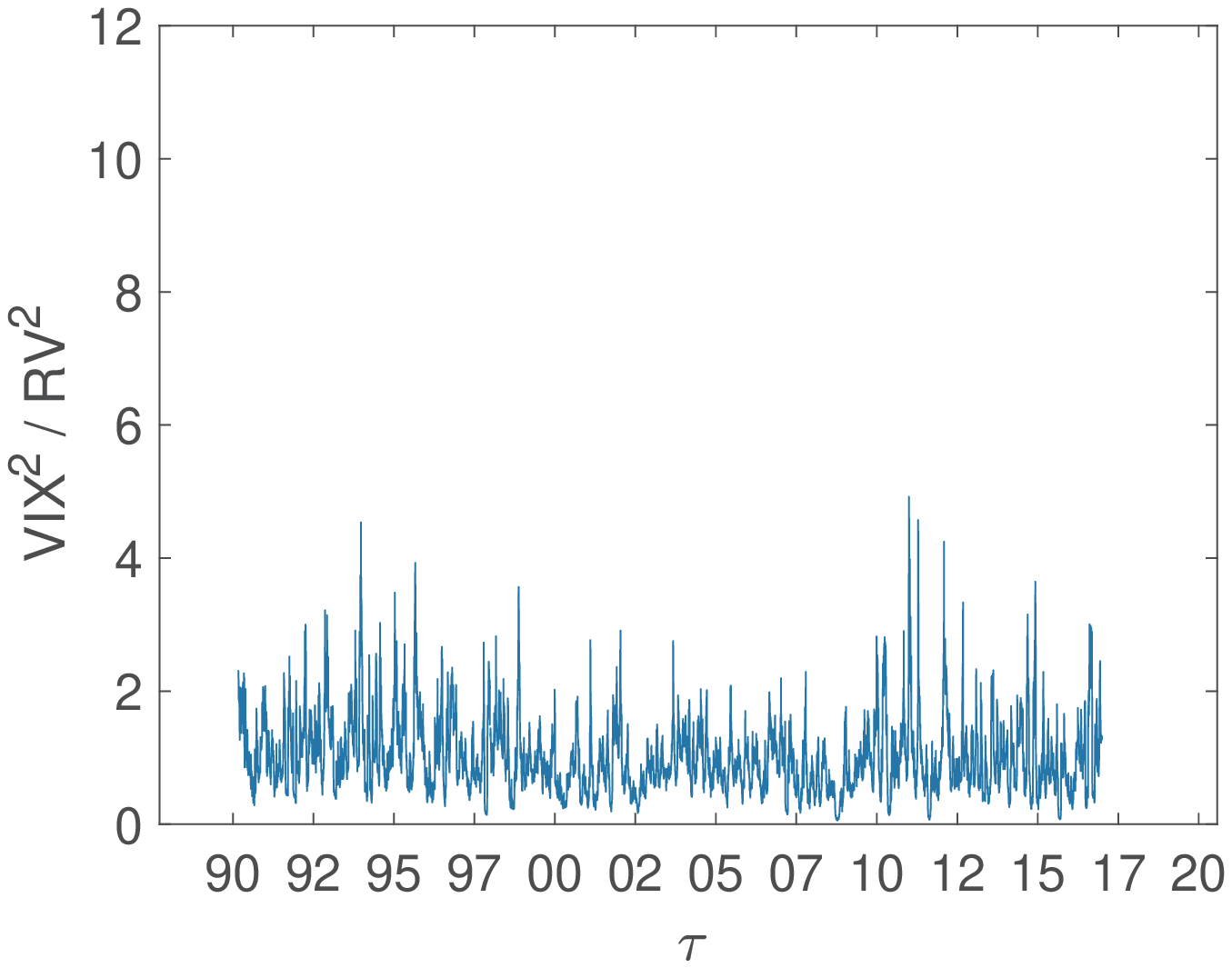}
\includegraphics[width = 0.49 \textwidth]{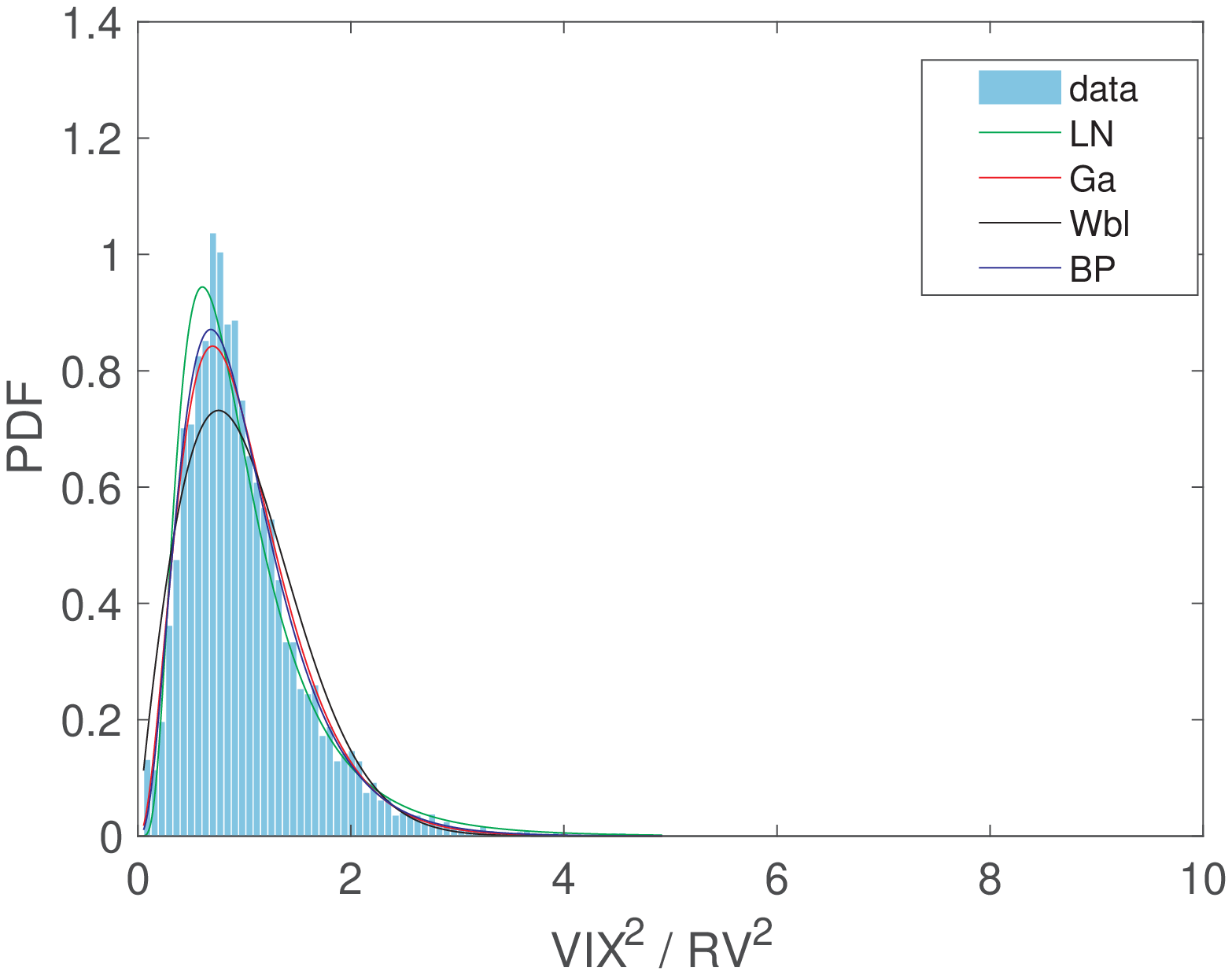}
\end{tabular}
\caption{$ \mathrm{VIX}^2 / \mathrm{RV}^2$, from Jan 2nd, 1990 to Dec 30th, 2016.}
\label{VIXOverRVListSVIX2OverRV21990nn}
\end{figure}

\begin{table}[!htb]
\caption{MLE results for ``$\mathrm{RV}^2 / \mathrm{VIX}^2$" and ``$\mathrm{VIX}^2 / \mathrm{RV}^2$"}
\label{MLESRV2OverVIX21990nn}
\begin{minipage}{0.5\textwidth}
\begin{center}
\begin{tabular}{ c c c} 
\multicolumn{2}{c}{} \\
\hline
             &       parameters &          KS test  \\
\hline
Normal & N(          1.0000,           0.9067) &           0.1940 \\
\hline
LogNormal & LN(         -0.2027,           0.5867) &           0.0446 \\
\hline
IGa & IGa(          3.3595,           2.3466) &           0.0246 \\
\hline
Gamma & Gamma(          2.6219,           0.3814) &           0.0978 \\
\hline
Weibull & Weibul(          1.1124,           1.4009) &           0.1224 \\
\hline
IG & IG(          1.0000,           2.3168) &           0.0607 \\
\hline
BP & BP(         27.2279,           3.8055,           0.1014) &           0.0198 \\
\hline
\hline
\end{tabular}
\end{center}
\end{minipage}
\begin{minipage}{.5\textwidth}
\begin{center}
\begin{tabular}{ c c c} 
\multicolumn{2}{c}{} \\
\hline
             &       parameters &          KS test  \\
\hline
Normal & N(          1.0000,           0.5626) &           0.0972 \\
\hline
LogNormal & LN(         -0.1562,           0.5867) &           0.0446 \\
\hline
IGa & IGa(          2.6219,           1.8314) &           0.0978 \\
\hline
Gamma & Gamma(          3.3595,           0.2977) &           0.0246 \\
\hline
Weibull & Weibul(          1.1306,           1.8882) &           0.0500 \\
\hline
IG & IG(          1.0000,           2.3168) &           0.0734 \\
\hline
BP & BP(          3.8055,          27.2279,           6.8913) &           0.0198 \\
\hline
\hline
\end{tabular}
\end{center}
  \end{minipage}

\end{table}

\clearpage

\begin{figure}[!htbp]
\centering
\begin{tabular}{cc}
\includegraphics[width = 0.49 \textwidth]{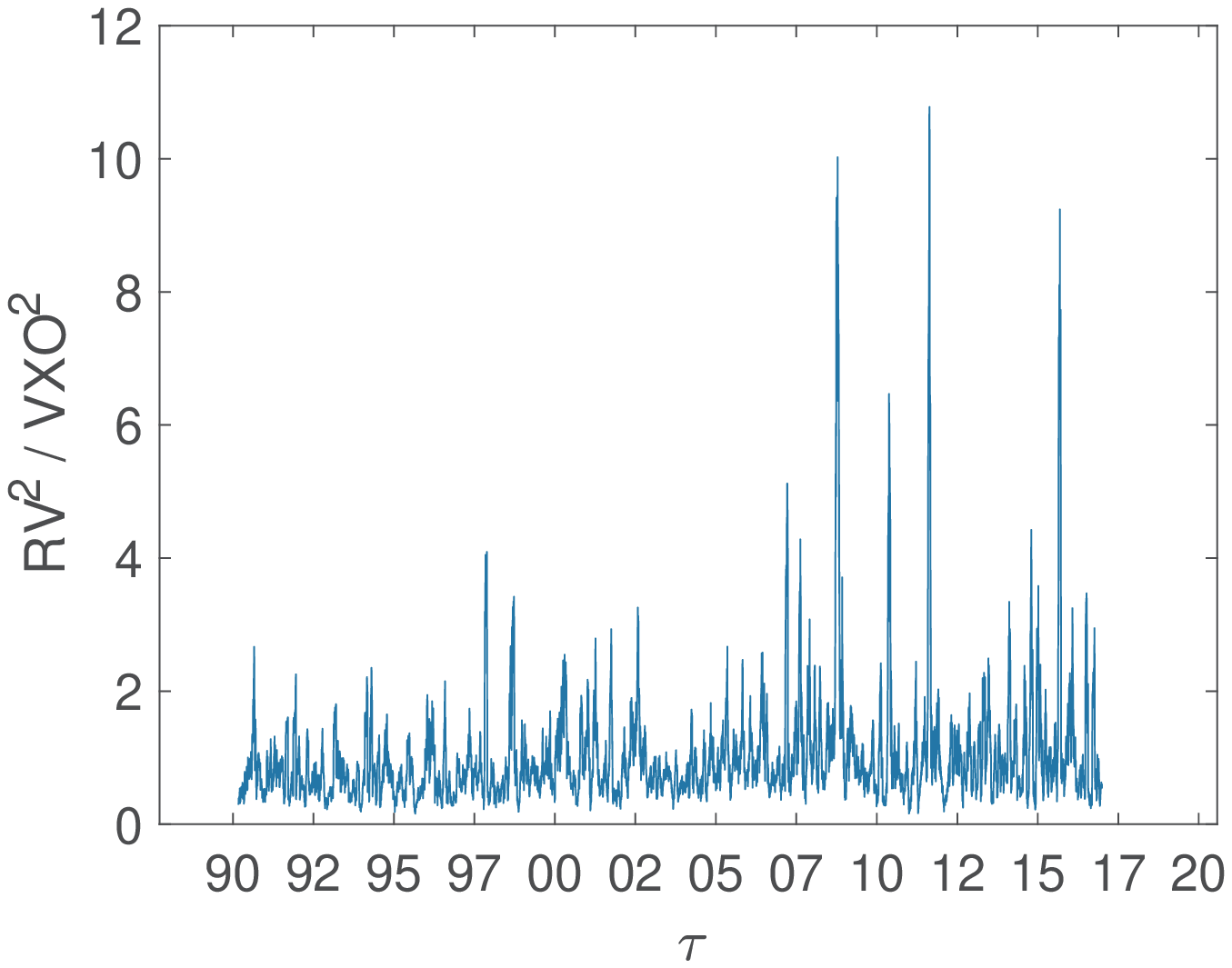}
\includegraphics[width = 0.49 \textwidth]{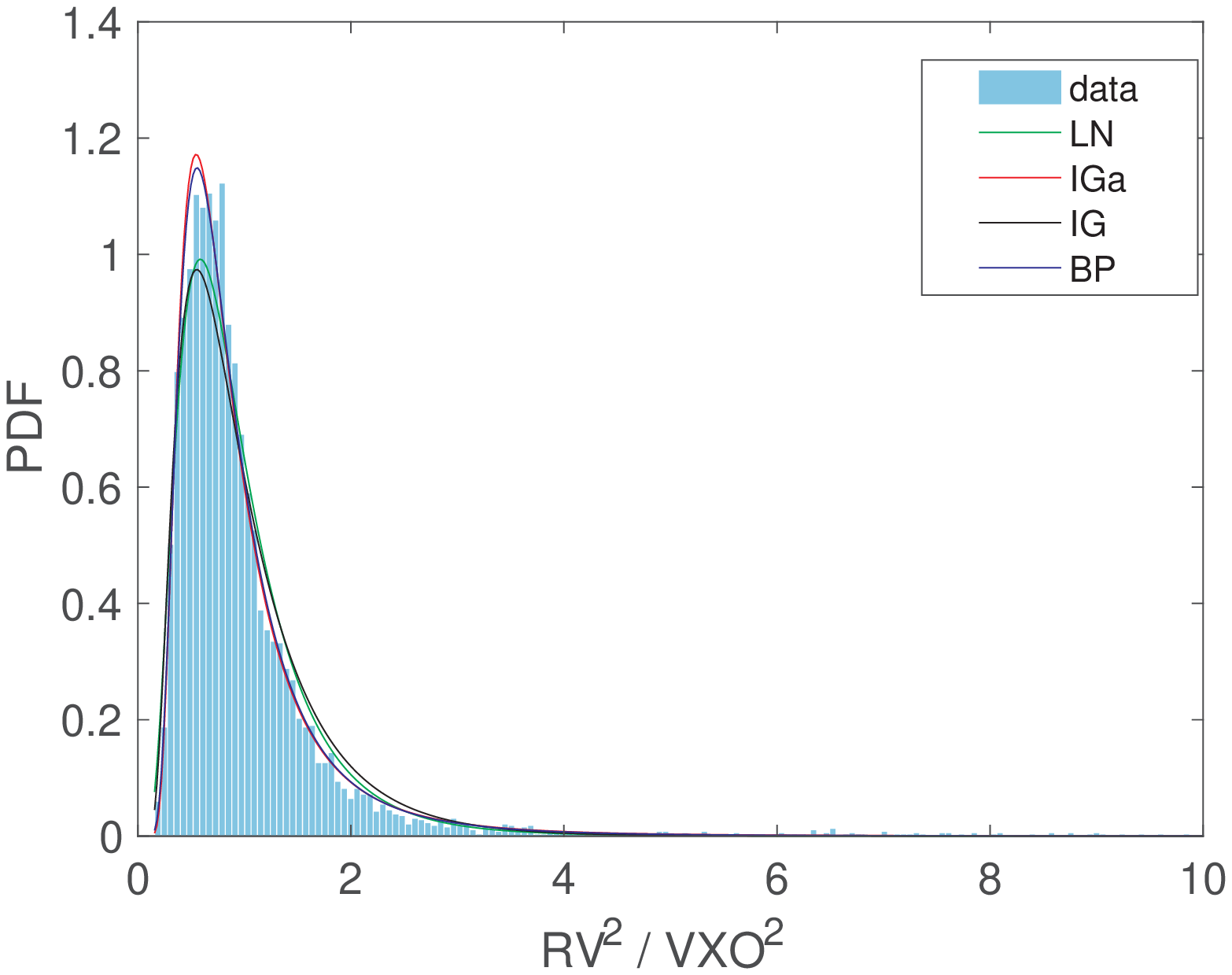}
\end{tabular}
\caption{$\mathrm{RV}^2 / \mathrm{VXO}^2$, Jan 2nd, 1990 to Dec 30th, 2016.}
\label{RVOverVXOListSRV2OverVXO22016}
\end{figure}

\begin{figure}[!htbp]
\centering
\begin{tabular}{cc}
\includegraphics[width = 0.49 \textwidth]{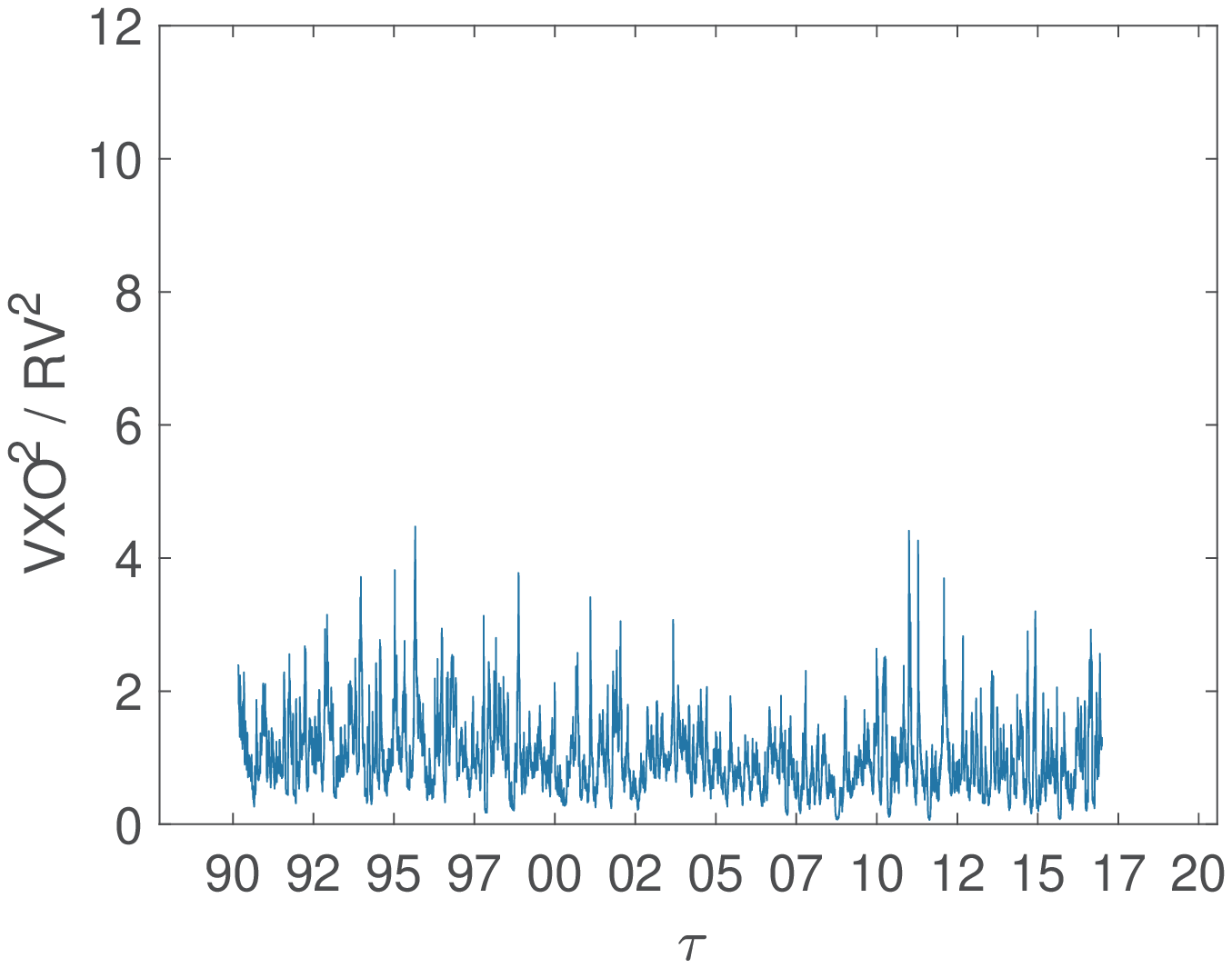}
\includegraphics[width = 0.49 \textwidth]{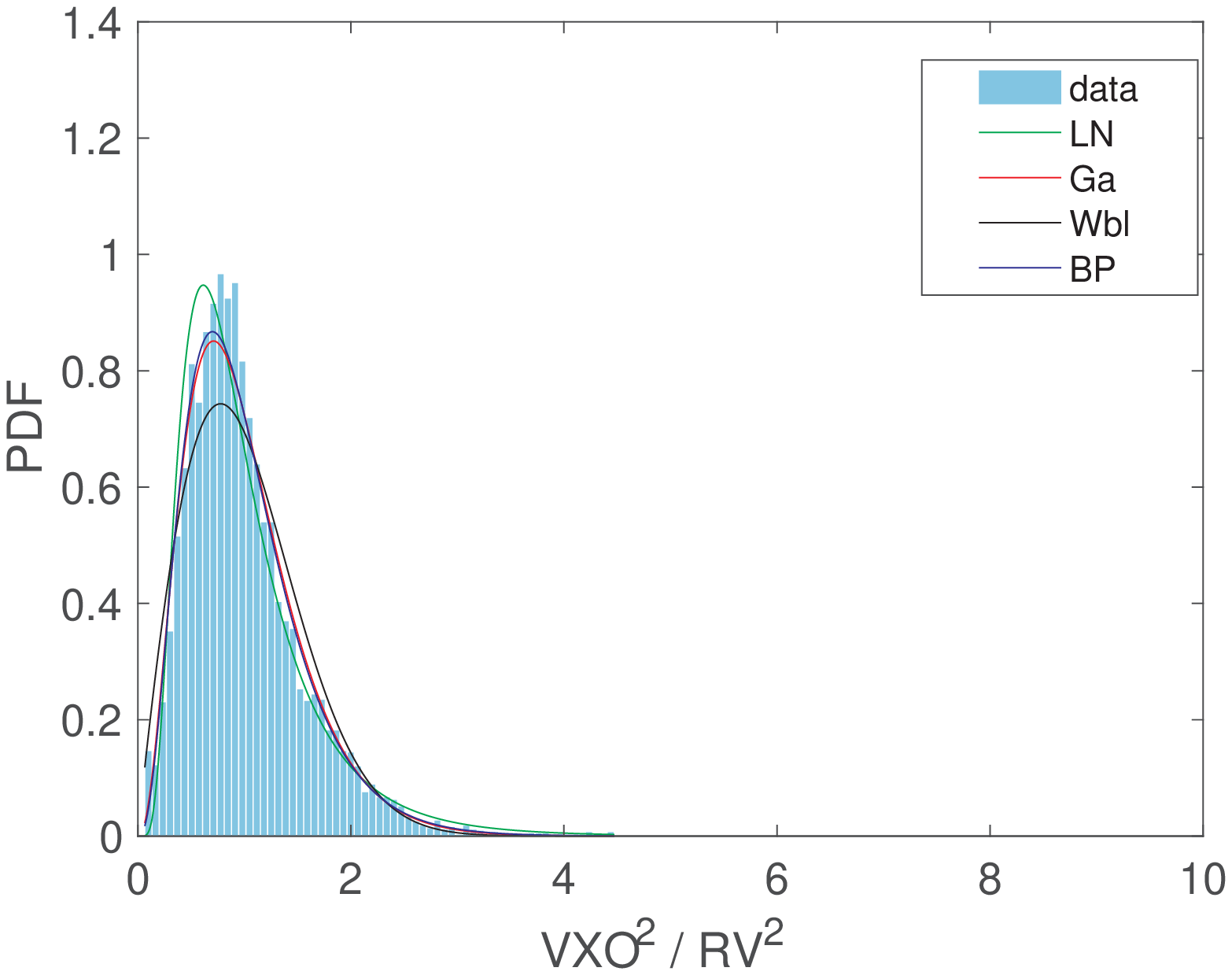}
\end{tabular}
\caption{$ \mathrm{VXO}^2 / \mathrm{RV}^2$, from Jan 2nd, 1990 to Dec 30th, 2016}
\label{VXOOverRVListSVXO2OverRV22016n}
\end{figure}

\begin{table}[!htb]
\caption{MLE results for ``$\mathrm{RV}^2 / \mathrm{VXO}^2$" and ``$\mathrm{VXO}^2 / \mathrm{RV}^2$"}
\label{MLESRV2OverVXO22016n}
\begin{minipage}{0.5\textwidth}
\begin{center}
\begin{tabular}{ c c c} 
\multicolumn{2}{c}{} \\
\hline
             &       parameters &          KS test  \\
\hline
Normal & N(          1.0000,           0.8747) &           0.1910 \\
\hline
LogNormal & LN(         -0.1973,           0.5795) &           0.0449 \\
\hline
IGa & IGa(          3.4629,           2.4438) &           0.0224 \\
\hline
Gamma & Gamma(          2.6897,           0.3718) &           0.0971 \\
\hline
Weibull & Weibul(          1.1150,           1.4256) &           0.1230 \\
\hline
IG & IG(          1.0000,           2.3981) &           0.0611 \\
\hline
BP & BP(         47.6001,           3.7157,           0.0563) &              0.0177 \\
\hline
\hline
\end{tabular}
\end{center}
\end{minipage}
\begin{minipage}{.5\textwidth}
\begin{center}
\begin{tabular}{ c c c} 
\multicolumn{2}{c}{} \\
\hline
             &       parameters &          KS test  \\
\hline
Normal & N(          1.0000,           0.5467) &           0.0925 \\
\hline
LogNormal & LN(         -0.1513,           0.5795) &           0.0449 \\
\hline
IGa & IGa(          2.6897,           1.8982) &           0.0971 \\
\hline
Gamma & Gamma(          3.4629,           0.2888) &           0.0224 \\
\hline
Weibull & Weibul(          1.1308,           1.9374) &           0.0499 \\
\hline
IG & IG(          1.0000,           2.3981) &           0.0729 \\
\hline
BP & BP(          3.7157,          47.6002,          12.5409) &           0.0177 \\
\hline
\hline
\end{tabular}
\end{center}
  \end{minipage}

\end{table}

\clearpage
\subsection{Preceding Month}

\begin{figure}[!htbp]
\centering
\begin{tabular}{cc}
\includegraphics[width = 0.49 \textwidth]{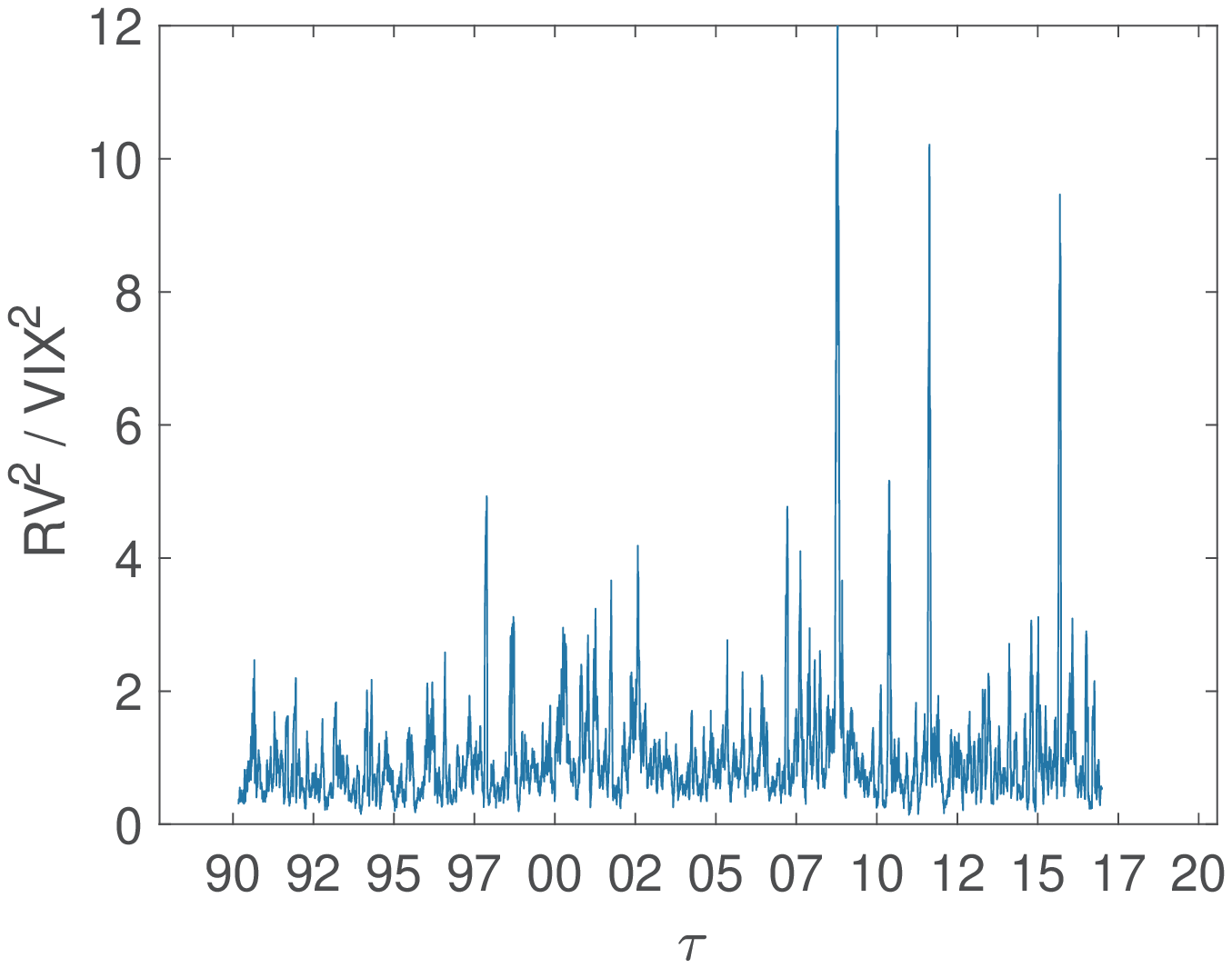}
\includegraphics[width = 0.49 \textwidth]{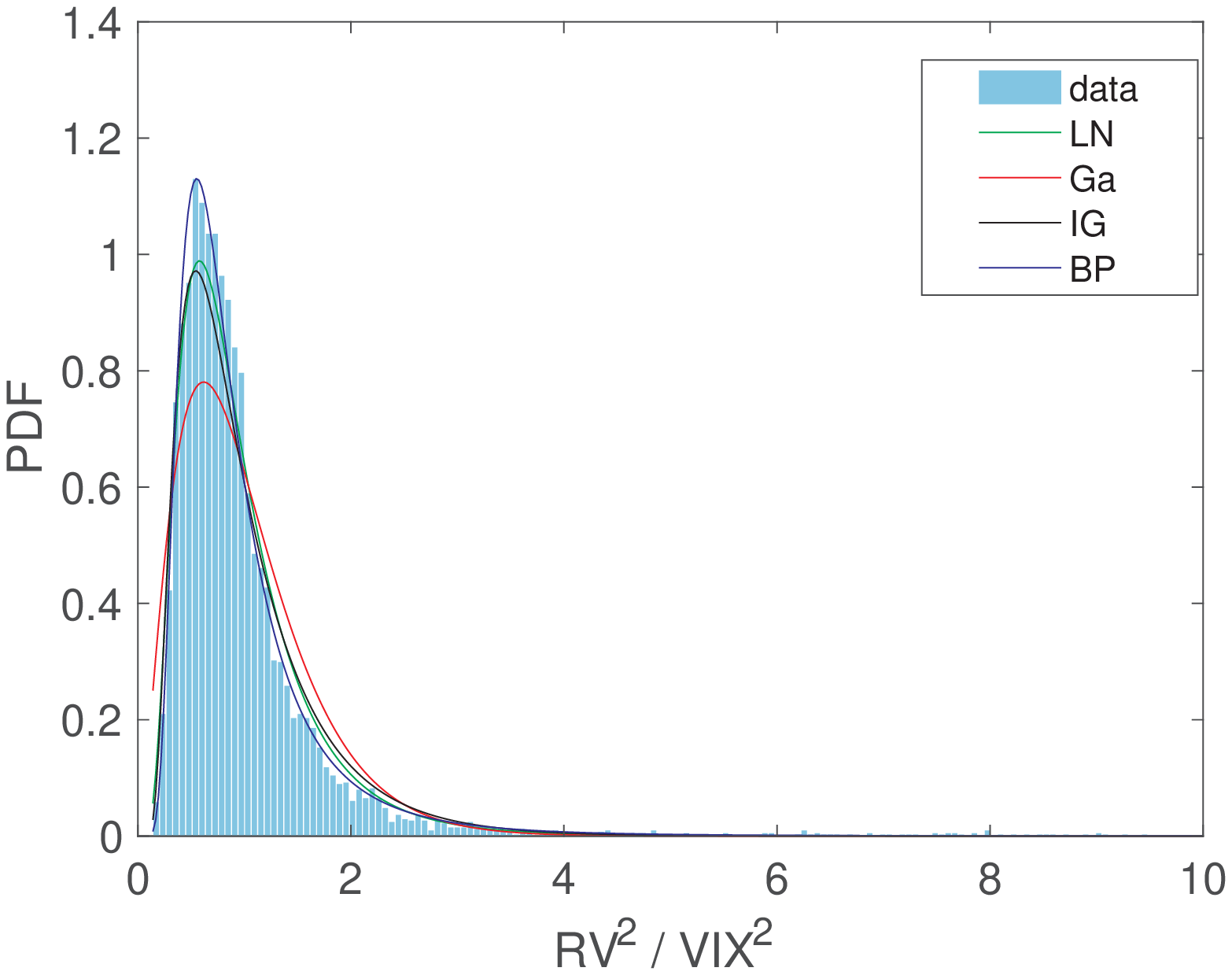}
\end{tabular}
\caption{$\mathrm{RV}^2 / \mathrm{VIX}^2$, from Jan 2nd, 1990 to Dec 30th, 2016.}
\label{RVOverVIXListSRV2OverVIX21990}
\end{figure}

\begin{figure}[!htbp]
\centering
\begin{tabular}{cc}
\includegraphics[width = 0.49 \textwidth]{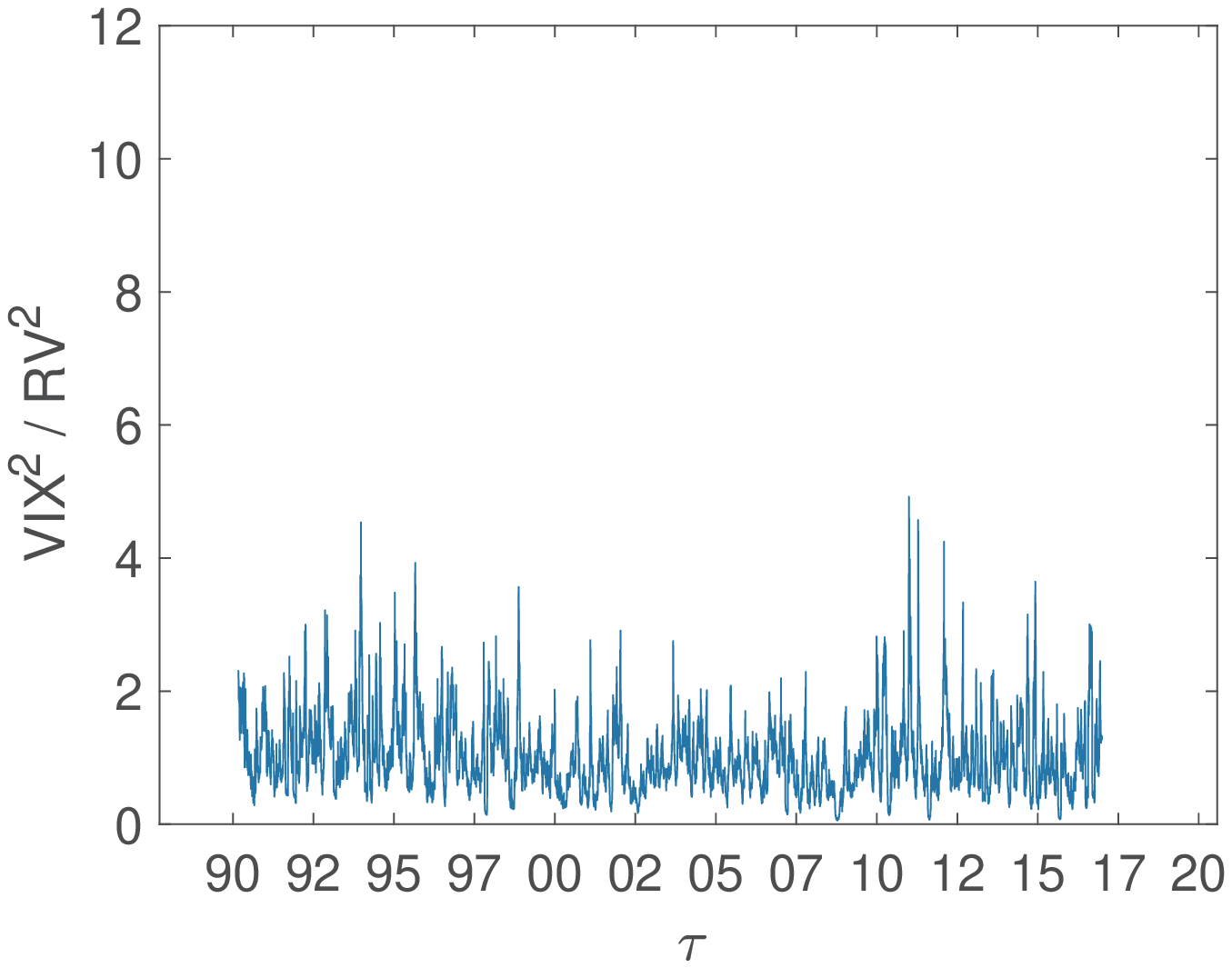}
\includegraphics[width = 0.49 \textwidth]{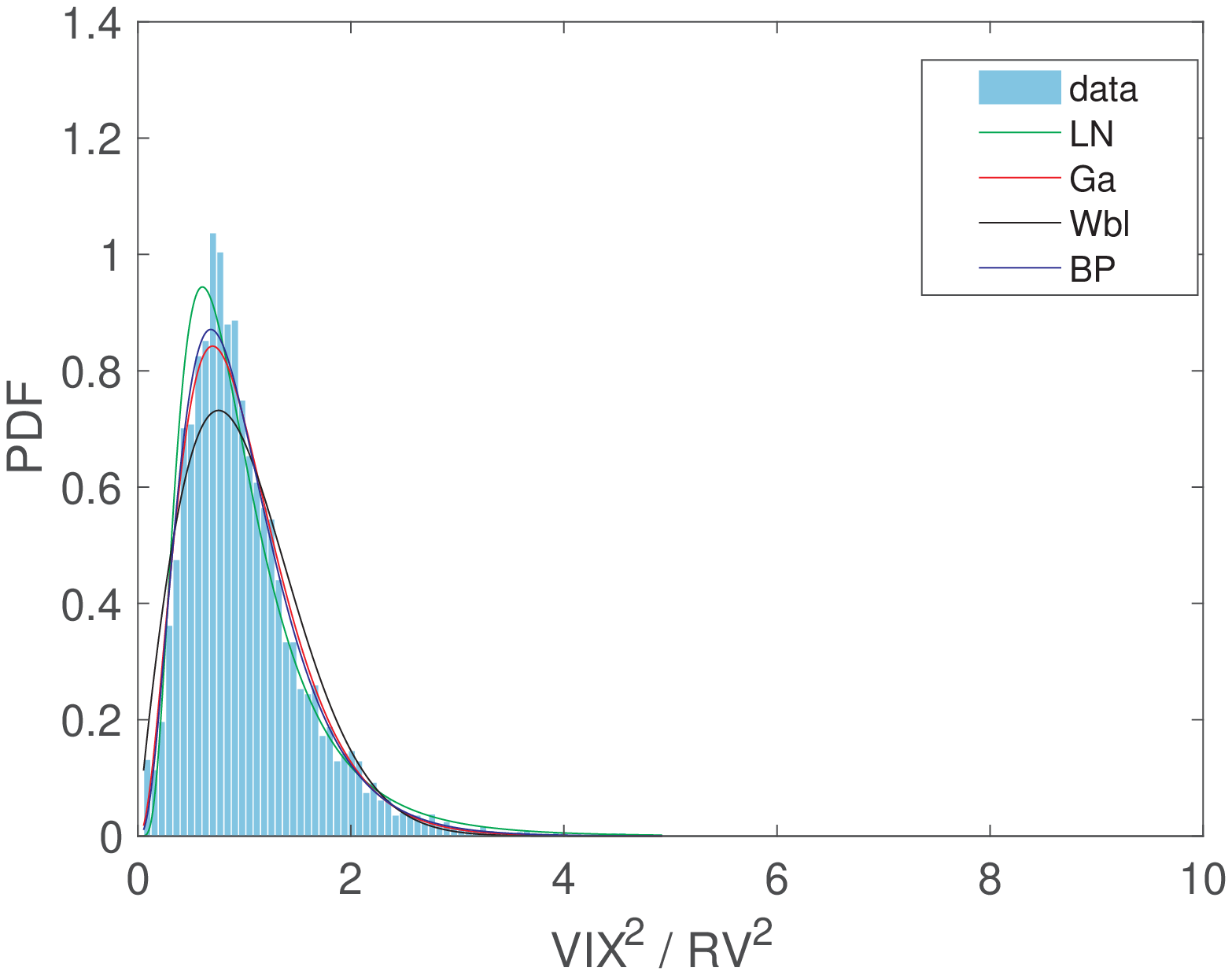}
\end{tabular}
\caption{$ \mathrm{VIX}^2 / \mathrm{RV}^2$, from Jan 2nd, 1990 to Dec 30th, 2016.}
\label{VIXOverRVListSVIX2OverRV21990nn}
\end{figure}

\begin{table}[!htb]
\caption{MLE results for ``$\mathrm{RV}^2 / \mathrm{VIX}^2$" and ``$\mathrm{VIX}^2 / \mathrm{RV}^2$"}
\label{MLESRV2OverVIX21990nn}
\begin{minipage}{0.5\textwidth}
\begin{center}
\begin{tabular}{ c c c} 
\multicolumn{2}{c}{} \\
\hline
             &       parameters &          KS test  \\
\hline
Normal & N(          1.0000,           0.4974) &           0.0992 \\
\hline
LogNormal & LN(         -0.1099,           0.4689) &           0.0147 \\
\hline
IGa & IGa(          4.6889,           3.7619) &           0.0431 \\
\hline
Gamma & Gamma(          4.7110,           0.2123) &           0.0381 \\
\hline
Weibull & Weibul(          1.1325,           2.1250) &           0.0672 \\
\hline
IG & IG(          1.0000,           4.0580) &           0.0215 \\
\hline
BP & BP(          9.2230,           9.9855,           0.9742) &           0.0117 \\
\hline
\hline
\end{tabular}
\end{center}
\end{minipage}
\begin{minipage}{.5\textwidth}
\begin{center}
\begin{tabular}{ c c c} 
\multicolumn{2}{c}{} \\
\hline
             &       parameters &          KS test  \\
\hline
Normal & N(          1.0000,           0.4999) &           0.1059 \\
\hline
LogNormal & LN(         -0.1104,           0.4689) &           0.0147 \\
\hline
IGa & IGa(          4.7110,           3.7796) &           0.0381 \\
\hline
Gamma & Gamma(          4.6889,           0.2133) &           0.0431 \\
\hline
Weibull & Weibul(          1.1329,           2.1186) &           0.0751 \\
\hline
IG & IG(          1.0000,           4.0580) &           0.0163 \\
\hline
BP & BP(          9.9855,           9.2230,           0.8236) &           0.0117 \\
\hline
\hline
\end{tabular}
\end{center}
  \end{minipage}

\end{table}

\clearpage

\begin{figure}[!htbp]
\centering
\begin{tabular}{cc}
\includegraphics[width = 0.49 \textwidth]{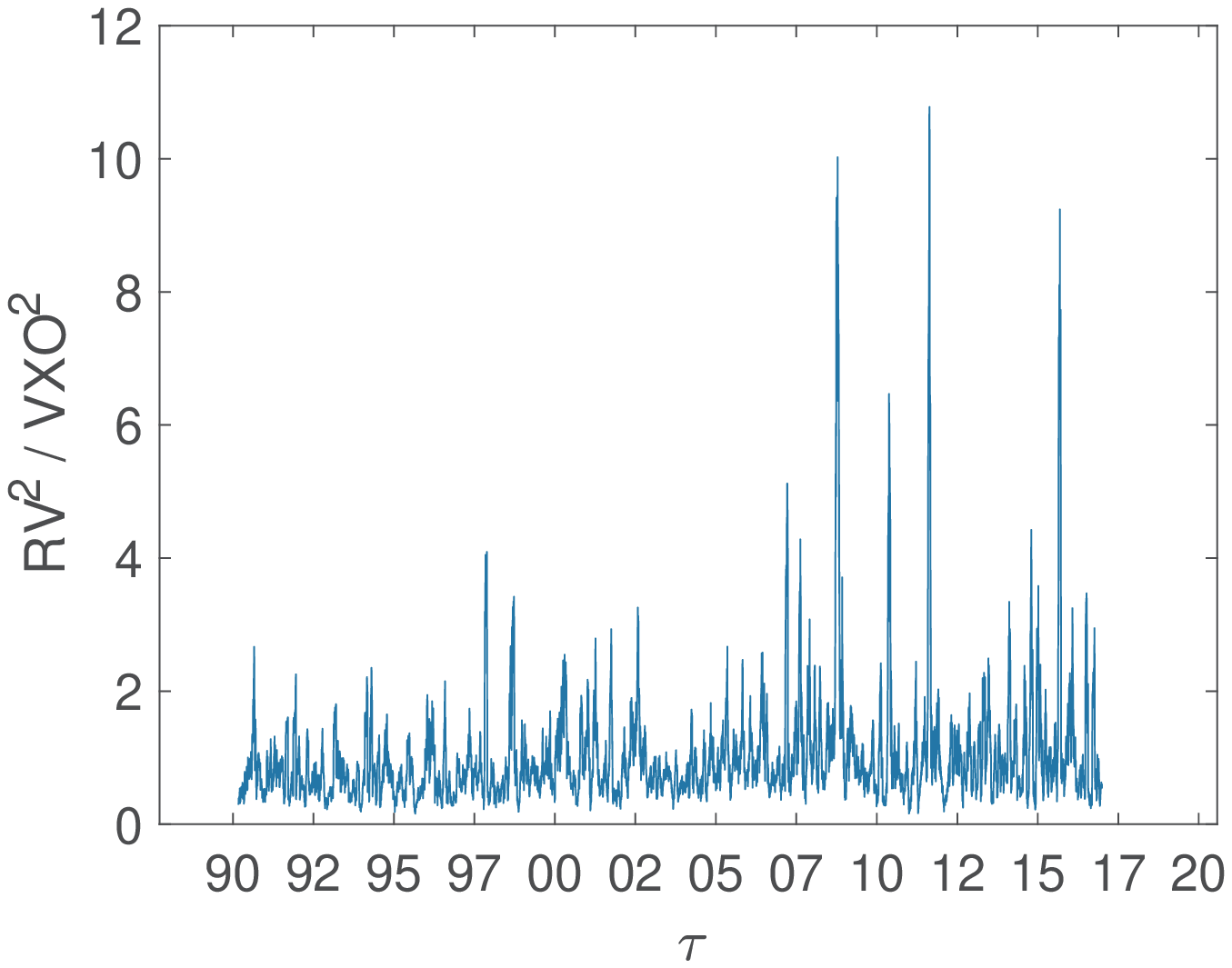}
\includegraphics[width = 0.49 \textwidth]{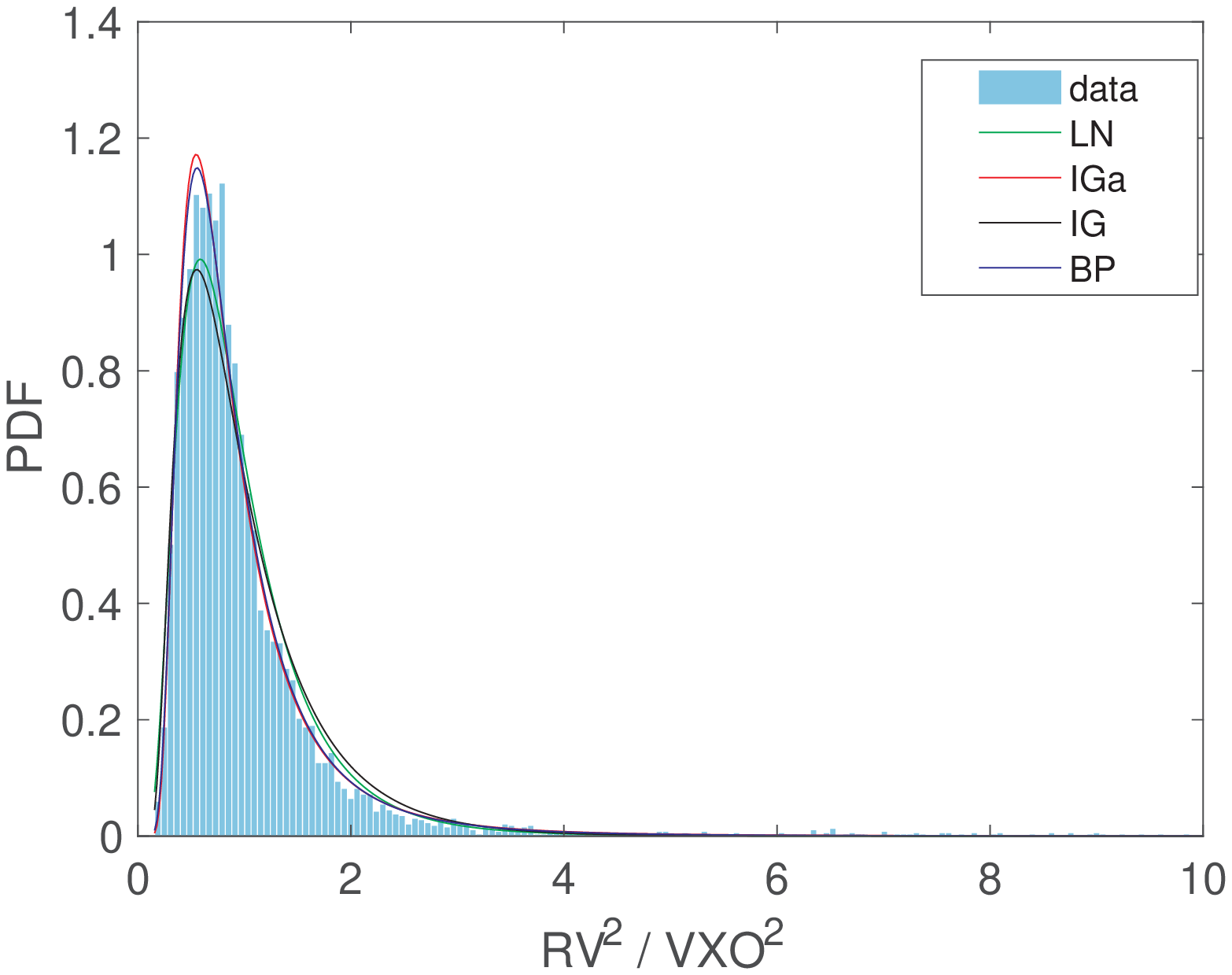}
\end{tabular}
\caption{$\mathrm{RV}^2 / \mathrm{VXO}^2$, Jan 2nd, 1990 to Dec 30th, 2016.}
\label{RVOverVXOListSRV2OverVXO22016}
\end{figure}

\begin{figure}[!htbp]
\centering
\begin{tabular}{cc}
\includegraphics[width = 0.49 \textwidth]{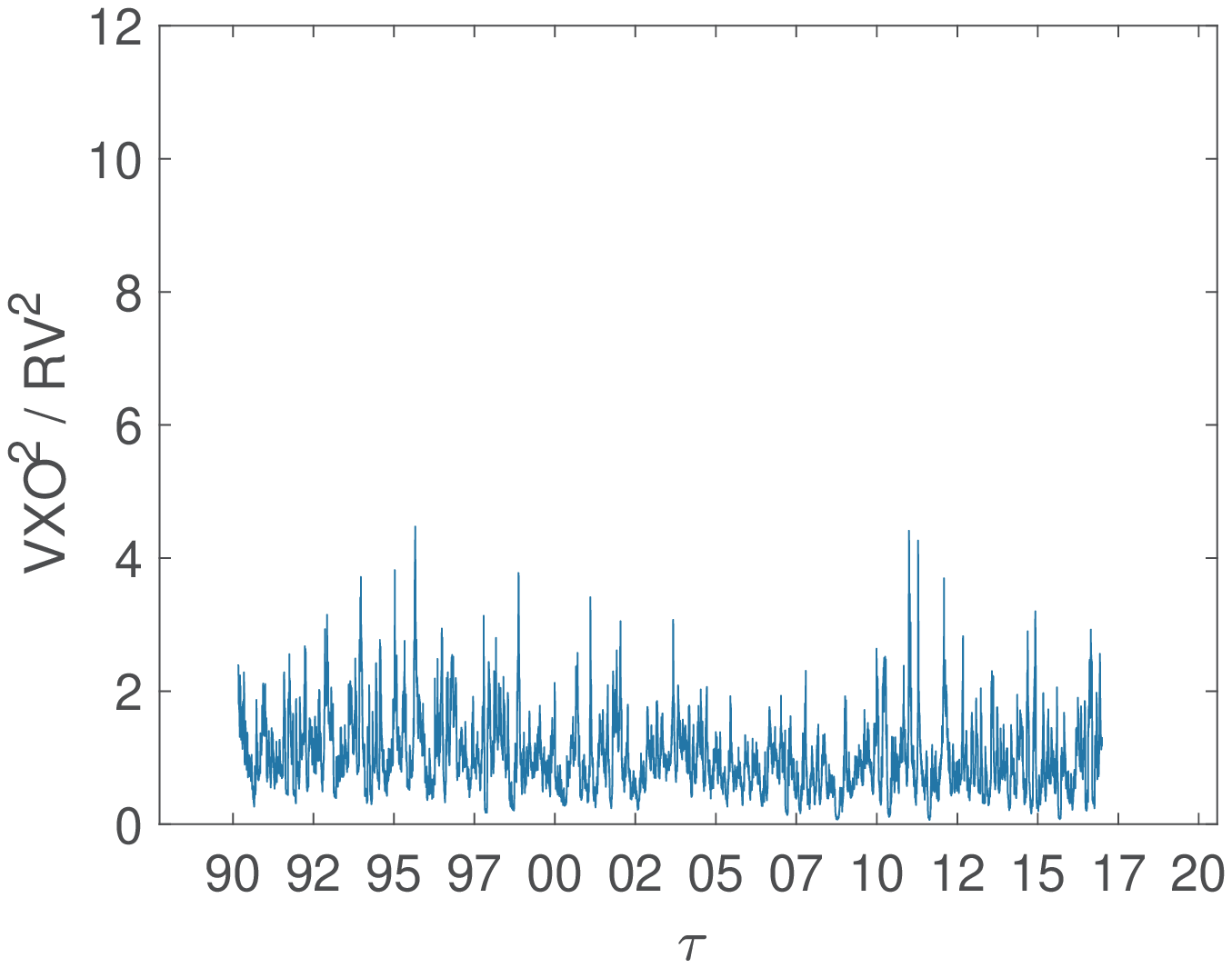}
\includegraphics[width = 0.49 \textwidth]{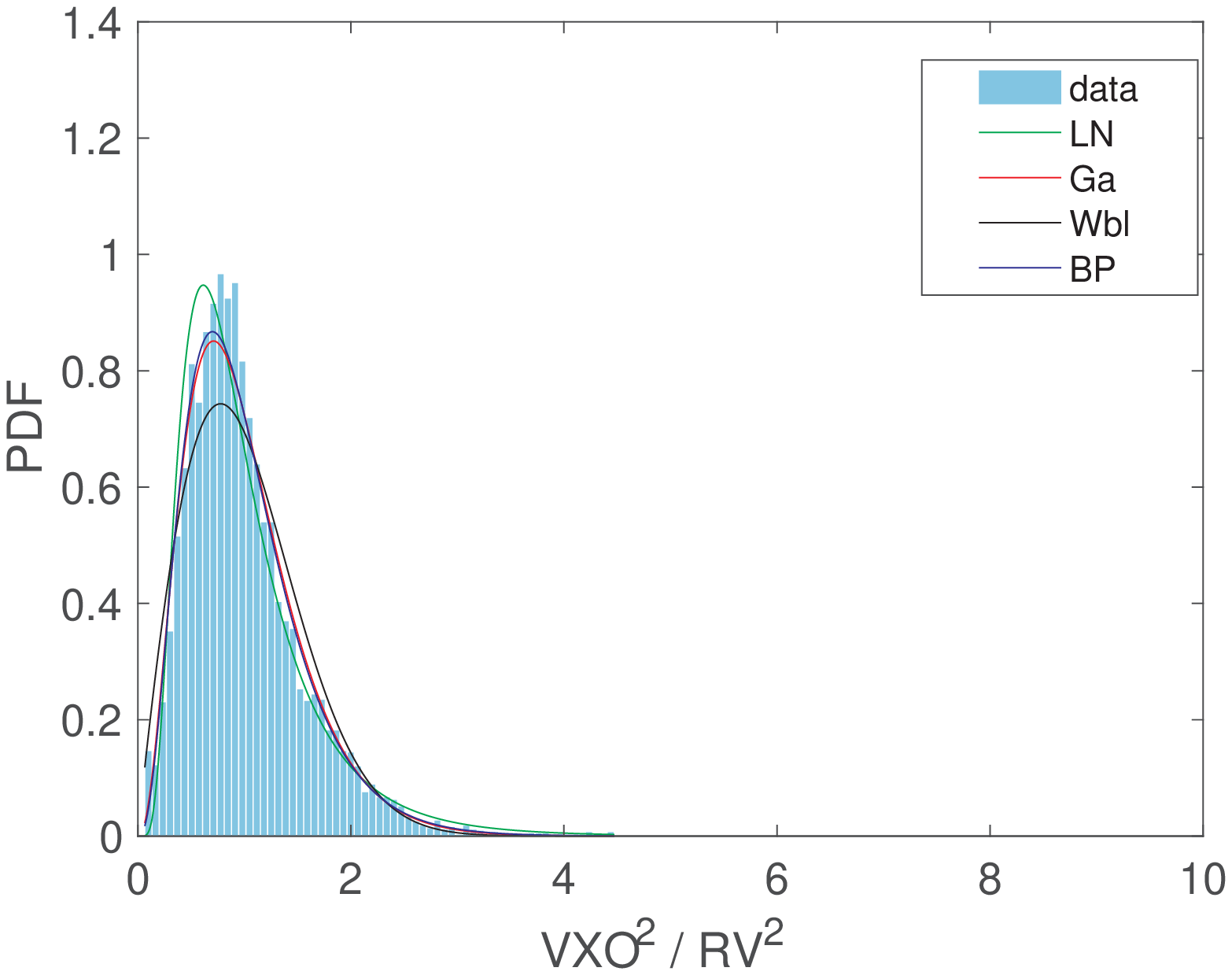}
\end{tabular}
\caption{$ \mathrm{VXO}^2 / \mathrm{RV}^2$, from Jan 2nd, 1990 to Dec 30th, 2016}
\label{VXOOverRVListSVXO2OverRV22016n}
\end{figure}

\begin{table}[!htb]
\caption{MLE results for ``$\mathrm{RV}^2 / \mathrm{VXO}^2$" and ``$\mathrm{VXO}^2 / \mathrm{RV}^2$"}
\label{MLESRV2OverVXO22016n}
\begin{minipage}{0.5\textwidth}
\begin{center}
\begin{tabular}{ c c c} 
\multicolumn{2}{c}{} \\
\hline
             &       parameters &          KS test  \\
\hline
Normal & N(          1.0000,           0.4915) &           0.1064 \\
\hline
LogNormal & LN(         -0.1041,           0.4539) &           0.0150 \\
\hline
IGa & IGa(          5.0351,           4.0948) &           0.0331 \\
\hline
Gamma & Gamma(          4.9618,           0.2015) &           0.0454 \\
\hline
Weibull & Weibul(          1.1316,           2.1383) &           0.0730 \\
\hline
IG & IG(          1.0000,           4.3548) &           0.0203 \\
\hline
BP & BP(         11.1694,           9.4027,           0.7520) &           0.0133 \\
\hline
\hline
\end{tabular}
\end{center}
\end{minipage}
\begin{minipage}{.5\textwidth}
\begin{center}
\begin{tabular}{ c c c} 
\multicolumn{2}{c}{} \\
\hline
             &       parameters &          KS test  \\
\hline
Normal & N(          1.0000,           0.4768) &           0.0933 \\
\hline
LogNormal & LN(         -0.1026,           0.4539) &           0.0150 \\
\hline
IGa & IGa(          4.9618,           4.0352) &           0.0454 \\
\hline
Gamma & Gamma(          5.0351,           0.1986) &           0.0331 \\
\hline
Weibull & Weibul(          1.1319,           2.2099) &           0.0689 \\
\hline
IG & IG(          1.0000,           4.3548) &           0.0212 \\
\hline
BP & BP(          9.4027,          11.1694,           1.0814) &           0.0133 \\
\hline
\hline
\end{tabular}
\end{center}
  \end{minipage}

\end{table}

\clearpage
\subsection{Ratio of Realized Variances of Two Adjacent Months}

\begin{figure}[!htbp]
\centering
\begin{tabular}{cc}
\includegraphics[width = 0.49 \textwidth]{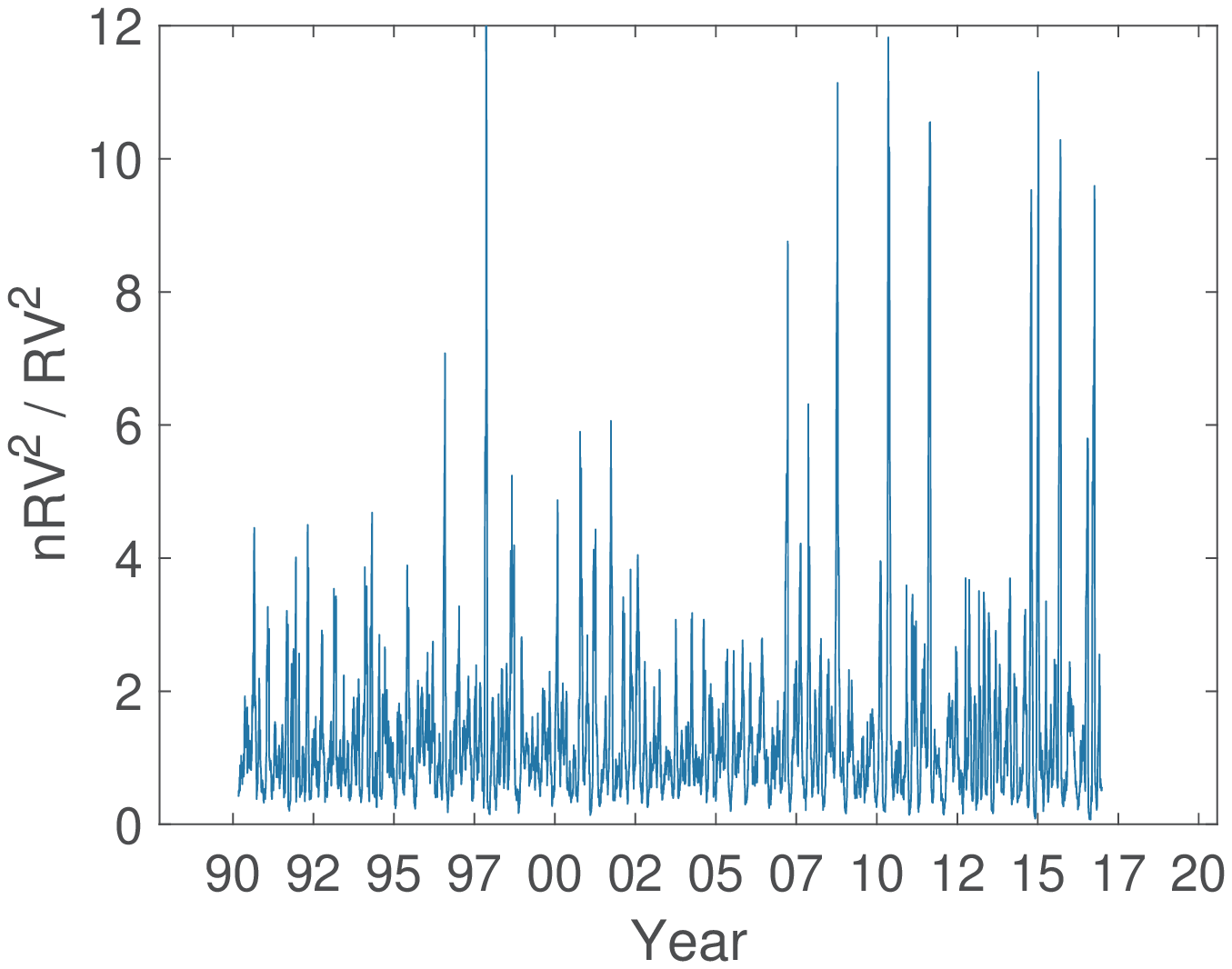}
\includegraphics[width = 0.49 \textwidth]{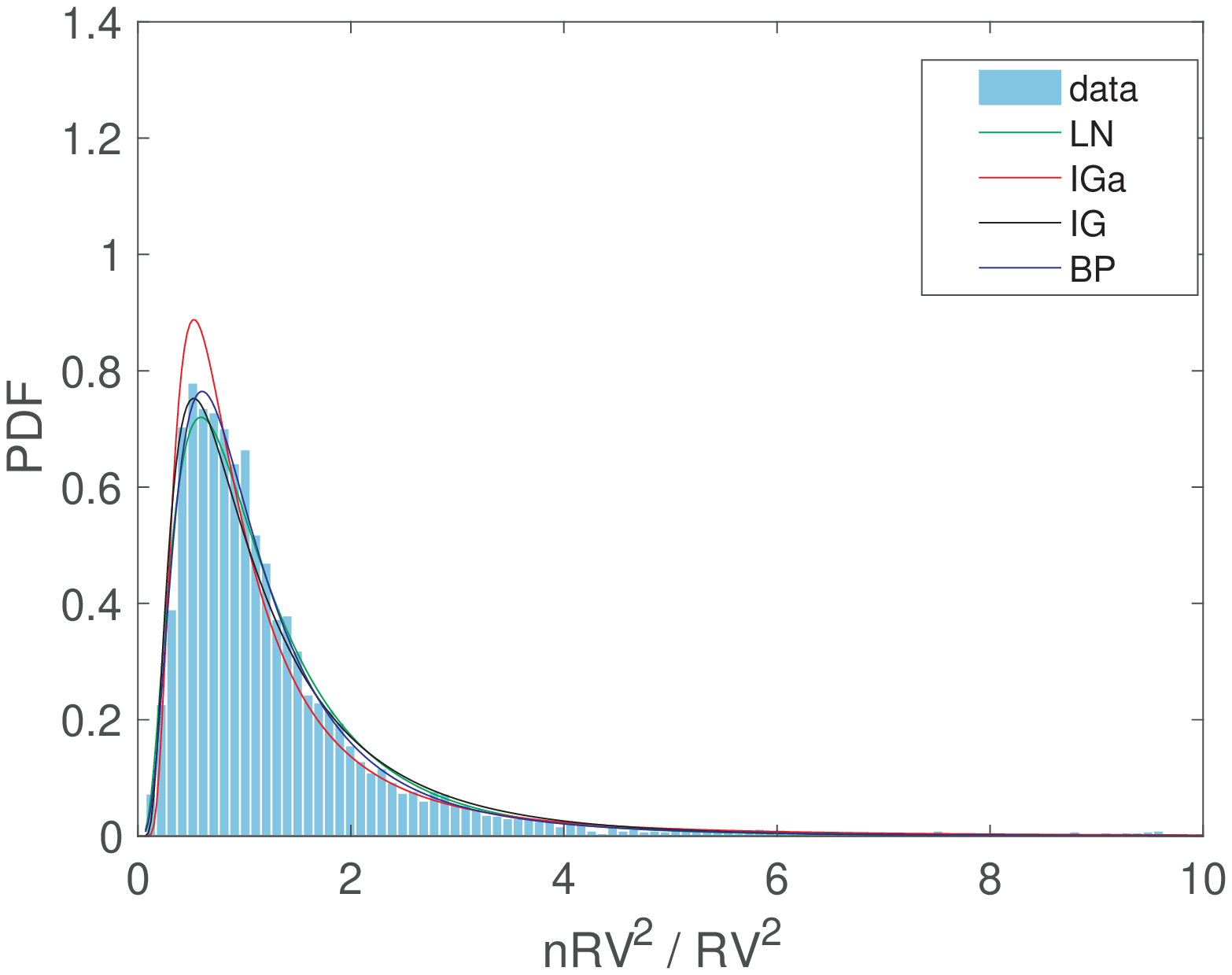}
\end{tabular}
\caption{Ratio of next-month realized variance to that of the preceding month, from Jan 2nd, 1990 to Dec 30th, 2016.}
\label{RVOverVIXListSRV2OverRV21990}
\end{figure}

\begin{figure}[!htbp]
\centering
\begin{tabular}{cc}
\includegraphics[width = 0.49 \textwidth]{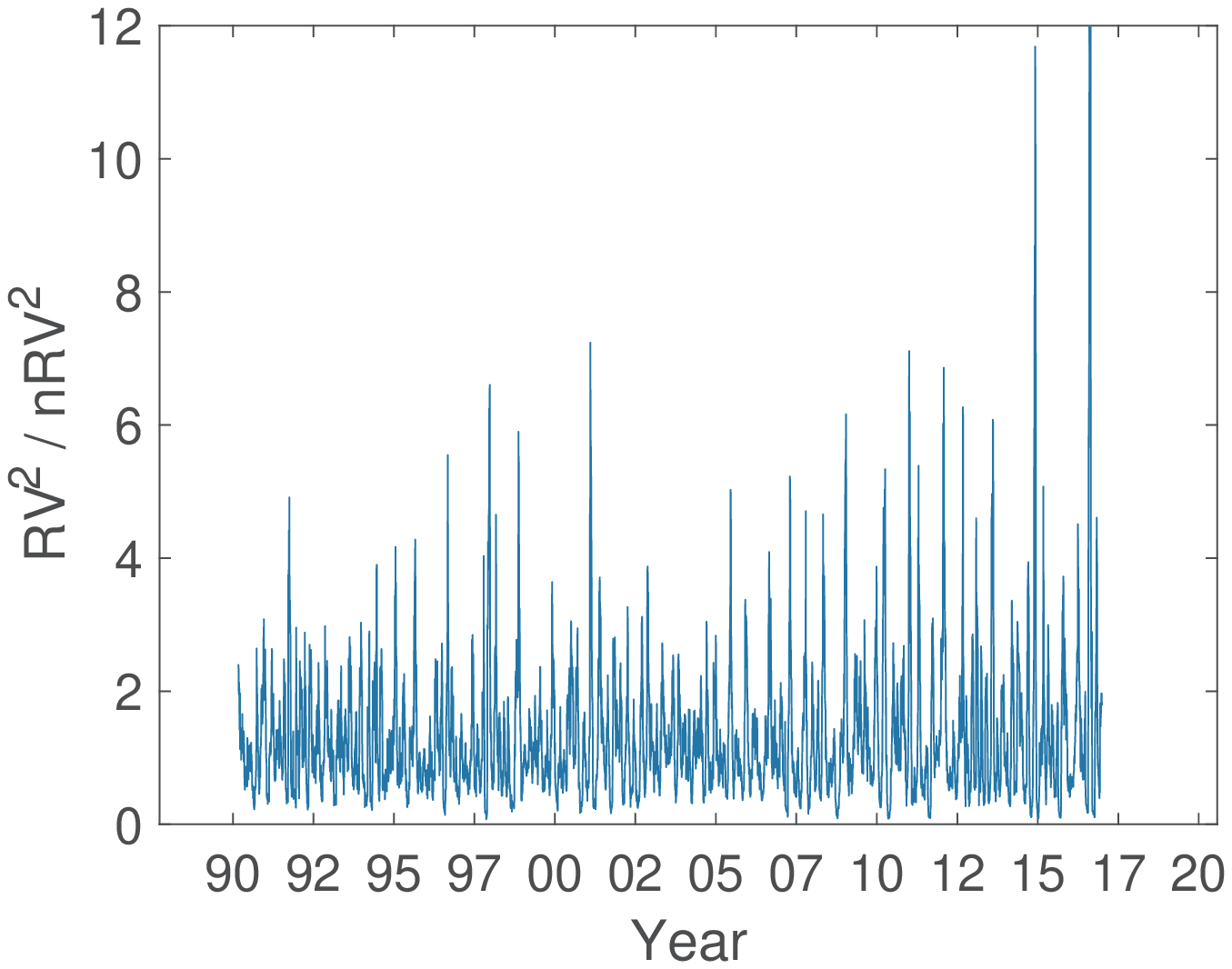}
\includegraphics[width = 0.49 \textwidth]{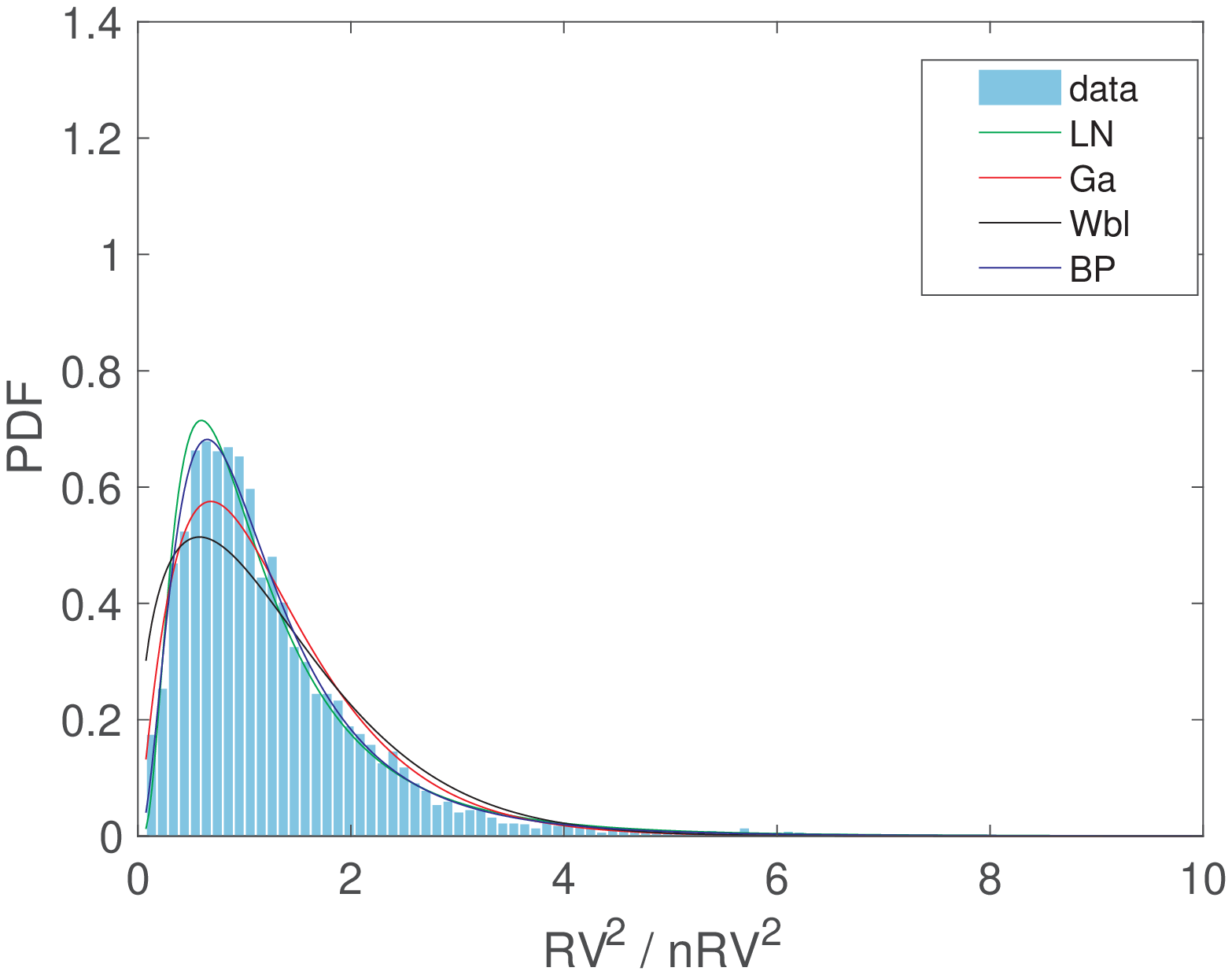}
\end{tabular}
\caption{Ratio of preceding-month realized variance to that of the following month, from Jan 2nd, 1990 to Dec 30th, 2016.}
\label{RVOverVIXListSRV2OverRV21990}
\end{figure}

\begin{table}[!htb]
\caption{MLE results for ``$\mathrm{nRV}^2 / \mathrm{RV}^2$" and ``$\mathrm{RV}^2 / \mathrm{nRV}^2$"}
\label{MLESRV2OverRV21990nn}
\begin{minipage}{0.5\textwidth}
\begin{center}
\begin{tabular}{ c c c} 
\multicolumn{2}{c}{} \\
\hline
             &       parameters &          KS test  \\
\hline
Normal & N(          1.3175,           1.2580) &           0.1809 \\
\hline
LogNormal & LN(         -0.0037,           0.7211) &           0.0244 \\
\hline
IGa & IGa(          2.1291,           1.6472) &           0.0472 \\
\hline
Gamma & Gamma(          1.9390,           0.6795) &           0.0801 \\
\hline
Weibull & Weibul(          1.4403,           1.2869) &           0.0922 \\
\hline
IG & IG(          1.3175,           1.8743) &           0.0340 \\
\hline
BP & BP(          5.8771,           3.4893,           0.5556) &           0.0123 \\
\hline
\hline
\end{tabular}
\end{center}
\end{minipage}
\begin{minipage}{.5\textwidth}
\begin{center}
\begin{tabular}{ c c c} 
\multicolumn{2}{c}{} \\
\hline
             &       parameters &          KS test  \\
\hline
Normal & N(          1.2925,           1.0777) &           0.1422 \\
\hline
LogNormal & LN(          0.0037,           0.7211) &           0.0244 \\
\hline
IGa & IGa(          1.9390,           1.4717) &           0.0801 \\
\hline
Gamma & Gamma(          2.1291,           0.6071) &           0.0472 \\
\hline
Weibull & Weibul(          1.4300,           1.3951) &           0.0608 \\
\hline
IG & IG(          1.2925,           1.8387) &           0.0513 \\
\hline
BP & BP(          3.4893,           5.8771,           1.7999) &           0.0123 \\
\hline
\hline
\end{tabular}
\end{center}
  \end{minipage}
\end{table}

\section{Correlation Analysis \label{Correlations}}
Tables \ref{PCCVIX} and \ref{PCCVXO} list Pearson correlation coefficients (PCC). Here "n" labels the "next" month, that is the month for which VIX and VXO were predicting the implied RV; "r" a "random" month and unlabeled RV is the one of the preceding month. All $RV^2$ are scaled, as explained earlier and in \cite{dashti2018implied}. Tables \ref{KSVIX} and \ref{KSVXO} list Kolmogorov-Smirnov (KS) statistic for comparison of the two plots.

\begin{table}[!htbp]
\centering
\caption{PCC VIX}
\label{PCCVIX}
\begin{tabular}{ccccc} 
& \multicolumn{4}{c}{}\\
\hline
             &       $RV^2$ & $nRV^2$ & $VIX^2$ & $rRV^2$  \\
\hline
$RV^2$ &1 & 0.70 & 0.88 & 0.0055 \\
\hline
$nRV^2$ &0.70 & 1 & 0.71 & 0.0025 \\
\hline
$VIX^2$ &0.88  & 0.71 & 1 & 0.003 \\
\hline
$rRV^2$ &0.0055  & 0.0025 & 0.003 & 1 \\
\hline
\hline
\end{tabular}
\end{table}

\begin{table}[!htbp]
\centering
\caption{PCC VXO}
\label{PCCVXO}
\begin{tabular}{ccccc} 
& \multicolumn{4}{c}{}\\
\hline
             &       $RV^2$ & $nRV^2$ & $VXO^2$ & $rRV^2$  \\
\hline
$RV^2$ &1 & 0.70 & 0.87 & 0.0015 \\
\hline
$nRV^2$ &0.70  & 1 & 0.72 & 0.004 \\
\hline
$VXO^2$ &0.87  & 0.72 & 1 & 0.002 \\
\hline
$rRV^2$ &0.0015  & 0.004 & 0.002 & 1 \\
\hline
\hline
\end{tabular}
\end{table}

\begin{table}[!htbp]
\centering
\caption{KS VIX}
\label{KSVIX}
\begin{tabular}{ccccccc} 
& \multicolumn{3}{c}{}\\
\hline
             &       $\frac{RV^2}{VIX^2}$ & $\frac{nRV^2}{VIX^2}$ & $\frac{RV^2}{nRV^2}$ & $\frac{rRV^2}{rVIX^2}$ & $\frac{rRV^2}{rRV^2}$& $\frac{nRV^2}{RV^2}$  \\ 
\hline
$\frac{RV^2}{VIX^2}$&0 & 0.056& 0.13& 0.20&0.26& -\\
\hline
$\frac{nRV^2}{VIX^2}$&0.056 & 0&- &0.18 & 0.23&0.13\\
\hline
$\frac{RV^2}{nRV^2}$  &0.13 &- & 0&  0.17&0.15&-\\          
\hline
$\frac{rRV^2}{rVIX^2}$ &0.20  & 0.18 &   0.17& 0 &0.063& 0.17\\
\hline
$\frac{rRV^2}{rRV^2}$ & 0.26 & 0.23 & 0.15 & 0.063& 0 &0.16 \\
\hline
$\frac{nRV^2}{RV^2}$  &- &0.13 & -& 0.17 &0.16&0 \\  
\hline
\hline
\end{tabular}
\end{table}

\begin{table}[!htbp]
\centering
\caption{KS VXO}
\label{KSVXO}
\begin{tabular}{ccccccc} 
& \multicolumn{3}{c}{}\\
\hline
             &       $\frac{RV^2}{VXO^2}$ & $\frac{nRV^2}{VXO^2}$ & $\frac{RV^2}{nRV^2}$ & $\frac{rRV^2}{rVXO^2}$ & $\frac{rRV^2}{rRV^2}$ & $\frac{nRV^2}{RV^2}$ \\ 
\hline
$\frac{RV^2}{VXO^2}$&0 & 0.063& 0.13& 0.22&0.26&- \\
\hline
$\frac{nRV^2}{VXO^2}$&0.063 & 0&- &0.19 & 0.23&0.16\\
\hline
$\frac{RV^2}{nRV^2}$  &0.13 &- & 0&  0.17&0.16&-\\          
\hline
$\frac{rRV^2}{rVXO^2}$ &0.22  & 0.19 &   0.17& 0 &0.057& 0.18\\
\hline
$\frac{rRV^2}{rRV^2}$ & 0.26 & 0.23 & 0.16  & 0.057& 0& 0.17 \\
\hline
$\frac{nRV^2}{RV^2}$  &- &0.16 & -&  0.18&0.17&0\\    
\hline
\hline
\end{tabular}
\end{table}

\clearpage

\section{Conclusions \label{Conclusions}}
Beta Prime distribution provides the best fit to the distributions of ratios of realized variance (squared realized volatility) to squared implied volatility indices VIX and VXO, as well as of ratio of realized variances of two consecutive months. 

For realized variance of the month for which volatility indices calculate implied realized variance, distributions have very slowly decaying fat tails. This indicates that volatility indices tends to underestimate future volatility, especially its large spikes. Conversely, probability of having very small ratios is suppressed due to a large power-law exponent. Comparing this to the ratio of realized variance of the preceding month to the following month, the latter has even longer tails, while the small ratios are significantly more populated. By both measures, VIX and VXO are better predictors of future realized volatility. 

For realized variance of the preceding month, the power law exponents for small ratios and for those in the tails are nearly identical, which reflects the fact that distributions of the ratio and its inverse are nearly identical and that the distribution of the inverse variable of Beta Prime is also Beta Prime.

Correlation and Kolmogorov-Smirnov statistics are in excellent agreement with empirical analysis of Section \ref{EO} and fitting in Section \ref{RV}. In a future work we will more closely identify the months whose ratios are responsible for the tails and the low-ratio regions.

\bibliography{mybib}

\end{document}